\documentclass[a4paper,11pt]{article}

\usepackage{jcappub}
\usepackage{amsmath}
\usepackage{amssymb}
\usepackage{latexsym}
\usepackage{graphicx}
\usepackage{hyperref}
\usepackage{mathrsfs}
\usepackage{color}
\usepackage{bm}
\usepackage[caption=false]{subfig}

\begin{document}

\title{Dodging the cosmic curvature to probe the constancy of the speed of light}
\author[a,b]{Rong-Gen Cai}

\author[a,b]{Zong-Kuan Guo}

\author[a,b]{Tao Yang}

\affiliation[a]{CAS Key Laboratory of Theoretical Physics, Institute of Theoretical Physics,
Chinese Academy of Sciences, P.O. Box 2735,
Beijing 100190, China}
\affiliation[b]{School of Physical Sciences, University of Chinese Academy of Sciences,
No.19A Yuquan Road, Beijing 100049, China}

\emailAdd{cairg@itp.ac.cn}
\emailAdd{guozk@itp.ac.cn}
\emailAdd{yangtao@itp.ac.cn}

\abstract{
We develop a new model-independent method to probe the constancy of the speed of light $c$. In our method, the degeneracy between the cosmic curvature and the speed of light can be eliminated, which makes the test more natural and general. Combining the independent observations of Hubble parameter $H(z)$ and luminosity distance $d_L(z)$, we use the model-independent smoothing technique, Gaussian processes, to reconstruct them and then detect variation of the speed of light. We find no signal of deviation from the present value of the speed of light $c_0$. Moreover, to demonstrate the improvement in probing the constancy of the speed of light from future experiments, we produce a series of simulated data. The Dark Energy Survey will be able to detect $\Delta c /c_0 \sim 1\%$ at $\sim 1.5\sigma$ confidence level and $\Delta c /c_0 \sim 2\%$ at $\sim 3\sigma$ confidence level. If the errors are reduced to one-tenth of the expected DES ones, it can detect a $\Delta c /c_0 \sim 0.1\%$ variation at $\sim 2\sigma$ confidence level.}

\maketitle
\section{Introduction \label{sec:introduction}}
The constancy of the speed of light is one of the most fundamental and recognized physical properties and plays a crucial role in basic physical laws such as Maxwell equations, special and general relativity and many others. Although many measurements of the speed of light have been performed since 1675 and ended with very accurate measurements today, some people argue that the speed of light $c$ may be dynamical and has been varying in the past. Such an idea has attracted a lot of interest recently because it can provide an alternative solution of horizon and the flatness problems in the non-inflationary cosmology. The theories of varying speed of light (VSL) have been considered~\cite{Barrow:1999is,Albrecht:1998ir,Barrow:1998he,Magueijo:2003gj}. However, a comparison of those theories with experimental data seems to be still missing.
Recently, V. Salzano {\it et al}. introduced a model-independent method to measure the speed of light through Baryon Acoustic Oscillations (BAO), and then it was followed by some newly updated results about its application to the forecast data~\cite{Salzano:2014lra}. In their paper, based on a spatially flat Friedmann-Robertson-Walker (FRW) universe they derived a relation among the speed of light, the angular diameter distance and the Hubble function,
\begin{equation}
{D_A(z_M)}\cdot H(z_M) = c(z_M),
\label{equa:DHc}
\end{equation}
where $z_M$ is the redshift at which the angular diameter ${D_A}(z_M)$ is maximal. They used the Gaussian Processes (GP) to reconstruct $D_A(z)$ and $H(z)$ and then found the redshift $z_M$. Finally, they used Eq.~(\ref{equa:DHc}) to obtain the speed of light at $z_M$ and probed the constancy of the speed of light. However, since $z_M$ is not covered by observational data sets, they only used mock data to test the accuracy of the method. More recently, they showed that Square Kilometer Array (SKA) can detect a $1\%$ variation in the speed of light at $3\sigma$ level~\cite{Salzano:2015mxk}. But smaller signals are hardly detected by already-planned future galaxy surveys.

 It seems that  the method has two drawbacks. The first is that the speed of light can be measured only at one redshift $z_M$, which is extremely limited by the redshift range covered by the data. The second and most important is that they ignore the cosmic curvature which is degenerated with the speed of light. Although the value of the cosmic curvature given by current data is very small, if we want to probe even smaller variation of the speed of light, it may be problematic. As a model-independent method, it is more general to include the cosmic curvature. In this paper, we develop a new and more general model-independent method which can dodge the cosmic curvature. We use the luminosity distance $d_L$ instead of the angular diameter distance $D_A$ to test the speed of light, so that we can use the large sample of real data such as cosmic chronometers (CC), BAO and SNeIa Union 2.1 or joint light-curve analysis (JLA) to probe the constancy of the speed of light over a wide range of redshift. We will also produce a series of mock data to test the accuracy of our method.

This paper is organized as follows. In Sec.~\ref{sec:tb}, we introduce the theoretical method to probe the constancy of the speed of light. In Sec.~\ref{sec:nulltest}, we apply GP method to the test using two independent data sets: CC+BAO and SNeIa Union 2.1 (and JLA). Furthermore it is followed by a series of simulated data tests. We give discussions and conclusions in Sec.~\ref{sec:discussion}.

\section{Method \label{sec:tb}}

For a FRW universe, given that the speed of light is a function of time, the line-element can be expressed as
\begin{equation}
d{s^2} =  - {c^2}(t)d{t^2} + {a^2}(t)\left[\frac{{d{r^2}}}{{1 - K{r^2}}} + {r^2}(d{\theta ^2} + {\sin ^2}\theta d{\phi ^2})\right],
\label{equa:ds}
\end{equation}
where $K = +1,-1,0$ corresponds to a closed, open and flat universe, respectively. Note that we write the metric with a preferred proper comoving time~\cite{Qi:2014zja,Moffat:2015uya}. We define $D(z)\equiv(H_0/c_0)(1+z)^{-1} d_L(z)$ as the normalized comoving distance, where $c_0$ is the value of the speed of light today, and $d_L(z)$ is the luminosity distance. Given that the speed of light is time dependent, we should reevaluate the luminosity distance from the definition:
\begin{equation}
d_L^2 \equiv \frac{{{L_s}}}{{4\pi F}},
\label{equa:dl2}
\end{equation}
where $L_s$ is the absolute luminosity of a source and $F$ is an observed flux. Setting $r=\sin \chi $ ($K=+1$), $r=\chi $ ($K=0$), and $r=\sinh \chi $ ($K=-1$) in Eq.~(\ref{equa:ds}), the 3-dimensional space line-element is expressed as
\begin{equation}
d{\sigma ^2} = d{\chi ^2} + {({f_K}(\chi ))^2}(d{\theta ^2} + {\sin ^2}\theta d{\phi ^2}),
\label{equa:dsigma}
\end{equation}
where
\begin{equation}
{f_K}(\chi ) = \begin{cases}
\sin \chi &(K =  + 1),\\
\chi &(K = 0),\\
\sinh \chi &(K =  - 1).
\end{cases}
\label{equa:fk}
\end{equation}
The function~(\ref{equa:fk}) can be written in a unified way:
\begin{equation}
{f_K}(\chi ) = \frac{1}{{\sqrt { - K} }}\sinh (\sqrt { - K} \chi ).
\label{equa:fku}
\end{equation}
We can easily check that $\chi$ is the comoving distance:
\begin{equation}
\chi  = \frac{1}{{{H_0}}}\int_0^z {\frac{{c(\tilde z)}}{{E(\tilde z)}}} d\tilde z.
\label{equa:chi}
\end{equation}
Note that the observed luminosity $L_0$ (detected at $\chi =0$ and $z=0$) is different from the absolute luminosity $L_s$ of the source (emitted at the comoving distance $\chi$ with the redshift $z$). The flux $F$ is defined by $F=L_0/S$, where $S=4\pi(f_K(\chi))^2$ is the area of a sphere at $z=0$. So the luminosity distance in Eq.~(\ref{equa:dl2}) yields $d_L^2 = (f_K(\chi))^2 L_s/L_0$. If we write the energy of light emitted at the time-interval $\Delta t_1$ to be $\Delta E_1$, then the absolute luminosity  $L_s=\Delta E_1/\Delta t_1$, and similarly the observed luminosity is given by $L_0=\Delta E_0/\Delta t_0$. Note that here the speed of light is time dependent, so $\Delta E_1/\Delta E_0 = \nu_1/\nu_0 = (c_1/\lambda_1)/(c_0/\lambda_0)=\hat c(z) (1+z)$, where $\hat c(z)\equiv c(z)/c_0$. Moreover, $\Delta t_0/\Delta t_1 = \nu_1/\nu_0 =\hat c(z) (1+z)$. Hence we find
\begin{equation}
d_L = \hat c(z) f_K(\chi)(1+z).
\label{equa:dl}
\end{equation}
Then combine the definition of $D(z)$ and Eqs.~(\ref{equa:fku})-(\ref{equa:dl}), we can derive
\begin{equation}
D(z) = \frac{\hat c(z)}{{\sqrt {{\Omega _K}} }}\sinh (\sqrt {{\Omega _K}} \int_0^z {\frac{{\hat c(\tilde z)}}{{E(\tilde z)}}} d\tilde z),
\label{equa:D}
\end{equation}
where $E(z)\equiv H(z)/H_0$, and $\Omega_K \equiv -K {{c_0}^2}/{H_0}^2$ is the dimensionless curvature density parameter at $z=0$.
Differentiating Eq.~(\ref{equa:D}) with the redshfit z,  we can get
\begin{equation}
{\Omega _K} = \frac{{{{[\hat c(z)D'(z) - \hat c'(z)D(z)]}^2}{E^2}(z) - {{\hat c}^6}(z)}}{{{D^2}(z){{\hat c}^4}(z)}}.
\label{equa:OmegaK}
\end{equation}
From Eq.~(\ref{equa:OmegaK}) we can see that there is degeneracy between VSL and curvature which is also pointed out in Ref.~\cite{Salzano:2015mxk}: the possible detection of a signal might be equally interpreted as ``VSL + null curvature'' or ``constant $c(z)$ + curvature''. Thus, to eliminate $\Omega _K$, using the second derivative of $D(z)$ we obtain
\begin{align}
0=&~~{\hat c^6}(z) + A(z){\hat c^2}(z) - B(z)\hat c(z)\hat c'(z) \nonumber\\
  &~~+ M(z)c{'^2}(z) - N(z)\hat c(z)\hat c''(z),
\label{equa:cz}
\end{align}
where
\begin{align}
A(z) = &~~\left[ {D''(z)E(z) + D'(z)E'(z)} \right]D(z)E(z)\nonumber\\
       &~~ - D{'^2}(z){E^2}(z),
\label{equa:Az}
\end{align}

\begin{equation}
B(z) = {E^2}(z)D(z)D'(z) + {D^2}(z)E(z)E'(z),
\label{equa:Bz}
\end{equation}
and
\begin{equation}
M(z) = 2N(z) = 2{D^2}(z){E^2}(z).
\label{equa:Mz}
\end{equation}
Eq.~(\ref{equa:cz}) is valid for any form of $c(z)$ which is the function of time no matter what the value of the cosmic curvature is. If we assume the constancy of the speed of light, thus $\hat c(z)=c(z)/c_0=1$, then Eq.~(\ref{equa:cz}) becomes
\begin{align}
T(z)&~~\equiv 1 + \left[ {D''(z)E(z) + D'(z)E'(z)} \right]D(z)E(z)\nonumber\\
    &~~- D{'^2}(z){E^2}(z)=0,
\label{equa:ctest}
\end{align}
 it always holds if the speed of light is constant. Any deviation of $T$  from  $0$  at redshift $z_*$ will indicate $c({z_*})$ is different from $c_0$. Thus we can test the speed of light at every redshift we want. This will make our test more flexible so that our method will not be limited to probing the constancy of the speed of light only at redshift $z_M$ as obtained in~\cite{Salzano:2014lra,Salzano:2015mxk}. On the contrary, we can choose the redshift where the quality of the data is better so that we can improve the precision of the test for the mock data.

In summary, we propose a null test Eq.~(\ref{equa:ctest}) to probe the constancy of the speed of light in a more general way, which is cosmological model-independent. The most important point is that we can dodge the cosmic curvature. Thus we need not assume the value of the cosmic curvature. It is emphasized that the only hypothesis is the FRW metric of the background. In addition, we would like to
mention here that once reconstruct $A(z)$, $B(z)$, and $M(z)$, we can obtain the dependence of the speed of light on the redshift through solving (\ref{equa:cz}). However, in this paper we pay attention on testing the constancy of the speed of light through (\ref{equa:ctest}).

\section{Null test using $H(z)$ and supernovae data \label{sec:nulltest}}
Given some  observational  data sets, it is crucial to use a model-independent method to reconstruct $E(z)$, $D(z)$, $E'(z)$, $D'(z)$ and $D''(z)$, in order to test the constancy of the speed of light following (\ref{equa:ctest}).  Note that here we  need also reconstruct the derivatives of the functions $E(z)$ and $D(z)$. Here we use the nonparametric approach Gaussian processes~\cite{Holsclaw:2010nb,Holsclaw:2010sk,Holsclaw:2011wi,Seikel:2012uu} to smooth the data and take the Gaussian processes in Python~\cite{Seikel:2012uu}. The detailed analysis and description of the GP method can be found in~\cite{Seikel:2012uu,Seikel:2013fda} or in our previous works~\cite{Cai:2015zoa,Cai:2015pia}. The analysis of the choice covariance kernel can be also found in~\cite{Seikel:2013fda}.

\subsection{Hubble rate data, Union 2.1 and JLA}

Following~\cite{Seikel:2012cs,Bilicki:2012ub} we proceed to an analysis based on observational Hubble data compiled from several sources, independent of SNeIa. We combine measurements of $H(z)$ obtained with two methods. One is cosmic chronometers, which are mainly passively evolving galaxies. There are $21$ data points compiled by Moresco et al.~\cite{Moresco:2012by,Moresco:2015cya}. The other is radial baryon acoustic oscillations from galaxy clustering in redshift surveys, which give $7$ data points of Hubble parameters from different experiments~\cite{Gaztanaga:2008xz,Chuang:2011fy,Blake:2012pj,Reid:2012sw}. We summarize the total $28$ data points in Table~\ref{tab:hdata}.


\begin{table}
\begin{centering}\begin{tabular}{cccc}
\hline
Index & $z$ & $H(z)$ & Refs. \tabularnewline
\hline
1 & $0.090$ & $69\pm 12$ & \cite{Moresco:2012by} \tabularnewline
\hline
2 & $0.170$ & $83\pm 8$ &\cite{Moresco:2012by} \tabularnewline
\hline
3 & $0.179$ & $75\pm 4$ &\cite{Moresco:2012by} \tabularnewline
\hline
4 & $0.199$ & $75\pm 5$ &\cite{Moresco:2012by} \tabularnewline
\hline
5 & $0.240$ & $79.69\pm 2.32$ & \cite{Gaztanaga:2008xz} \tabularnewline
\hline
6 & $0.270$ & $77\pm 14$ &\cite{Moresco:2012by} \tabularnewline
\hline
7 & $0.350$ & $82.1\pm 4.9$ & \cite{Chuang:2011fy} \tabularnewline
\hline
8 & $0.352$ & $83\pm 14$ &\cite{Moresco:2012by} \tabularnewline
\hline
9 & $0.400$ & $95\pm 17$ &\cite{Moresco:2012by} \tabularnewline
\hline
10 & $0.430$ & $86.45\pm 3.27$ & \cite{Gaztanaga:2008xz} \tabularnewline
\hline
11 & $0.440$ & $82.6\pm 7.8$ &  \cite{Blake:2012pj} \tabularnewline
\hline
12 & $0.480$ & $97\pm 62$ & \cite{Moresco:2012by} \tabularnewline
\hline
13 & $0.570$ & $92.4\pm 4.5$ & \cite{Reid:2012sw} \tabularnewline
\hline
14 & $0.593$ & $104\pm 13$ & \cite{Moresco:2012by} \tabularnewline
\hline
15 & $0.600$ & $87.9\pm 6.1$ & \cite{Blake:2012pj} \tabularnewline
\hline
16 & $0.680$ & $92\pm 8$ & \cite{Moresco:2012by} \tabularnewline
\hline
17 & $0.730$ & $97.3\pm 7$ &  \cite{Blake:2012pj} \tabularnewline
\hline
18 & $0.781$ & $105\pm 12$ & \cite{Moresco:2012by} \tabularnewline
\hline
19 & $0.875$ & $125\pm 17$ & \cite{Moresco:2012by} \tabularnewline
\hline
20 & $0.880$ & $90\pm 40$ & \cite{Moresco:2012by} \tabularnewline
\hline
21 & $0.900$ & $117\pm 23$ & \cite{Moresco:2012by} \tabularnewline
\hline
22 & $1.037$ & $154\pm 20$ & \cite{Moresco:2012by} \tabularnewline
\hline
23 & $1.300$ & $168\pm 17$ & \cite{Moresco:2012by} \tabularnewline
\hline
24 & $1.363$ & $160\pm 33.6$ & \cite{Moresco:2015cya} \tabularnewline
\hline
25 & $1.430$ & $177\pm 18$ &\cite{Moresco:2012by} \tabularnewline
\hline
26 & $1.530$ & $140\pm 14$ & \cite{Moresco:2012by} \tabularnewline
\hline
27 & $1.750$ & $202\pm 40$ & \cite{Moresco:2012by} \tabularnewline
\hline
28 & $1.965$ & $186.5\pm 50.4$ & \cite{Moresco:2015cya} \tabularnewline
\hline
\end{tabular}\par\end{centering}
\caption{$H(z)$ measurements from different surveys using passively evolving galaxies and radial BAO.
\label{tab:hdata}}
\end{table}

\begin{figure}
\centering
\includegraphics[width=0.45\textwidth]{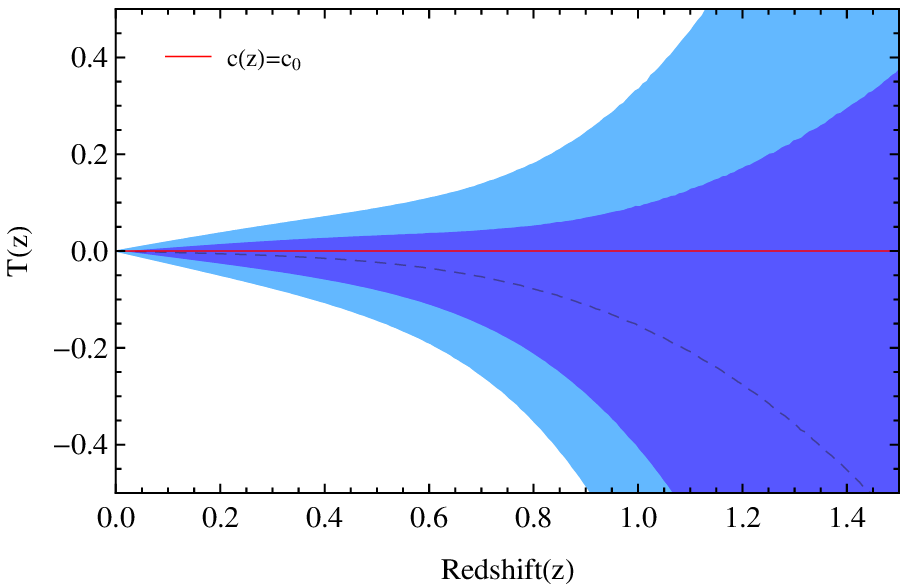}\quad
\includegraphics[width=0.45\textwidth]{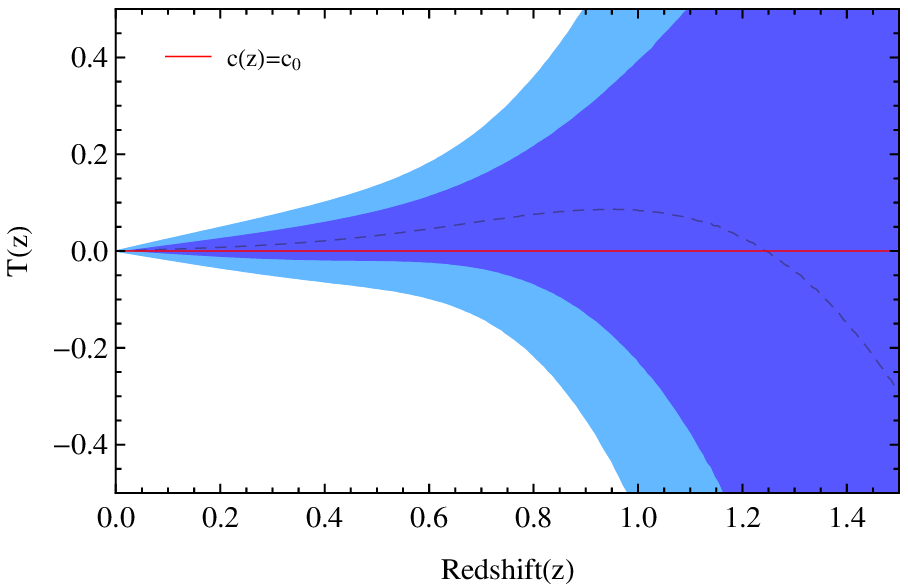}
\caption{Reconstruction of $T(z)$ from CC+BAO with Union 2.1 (Left) and JLA (Right). The shaded blue regions are the $68\%$ and $95\%$ C.L. of the reconstruction. The red line corresponds to $c_0$.}
\label{fig:realc}
\end{figure}

To reconstruct $D(z)$, we use two different data, first we adopt the SNeIa Union 2.1 data sets~\cite{Suzuki:2011hu}, which contain $580$ SNeIa data. Then we employ the most recent SNeIa catalog available: the JLA~\cite{Betoule:2014frx}. We use the binned JLA compilation which shows the same trend as using the full catalog itself. For Union 2.1 and the JLA, We have included the covariance matrix with both the statistical and systematic uncertainties. We transform the distance modulus $m-M$ given in the data set to $D$ using
\begin{equation}
m-M+5\log\left[\frac{H_0}{c_0}\right]-25=5\log\left[(1+z)D\right].
\end{equation}
We apply the reconstructions of $E(z)$, $D(z)$ and their derivatives to the null test in Eq.~(\ref{equa:ctest}). The result is shown in Fig.~\ref{fig:realc}.

As expected, from Fig.~\ref{fig:realc} we can see that for both of the Union 2.1 and the JLA, $T(z)=0$ lies within $1\sigma$ C.L. of the reconstructed region, which indicates that using current observational data there is no signal of variation of $c(z)$.

\begin{figure}
\centering
\subfloat{
\includegraphics[width=0.25\textwidth]{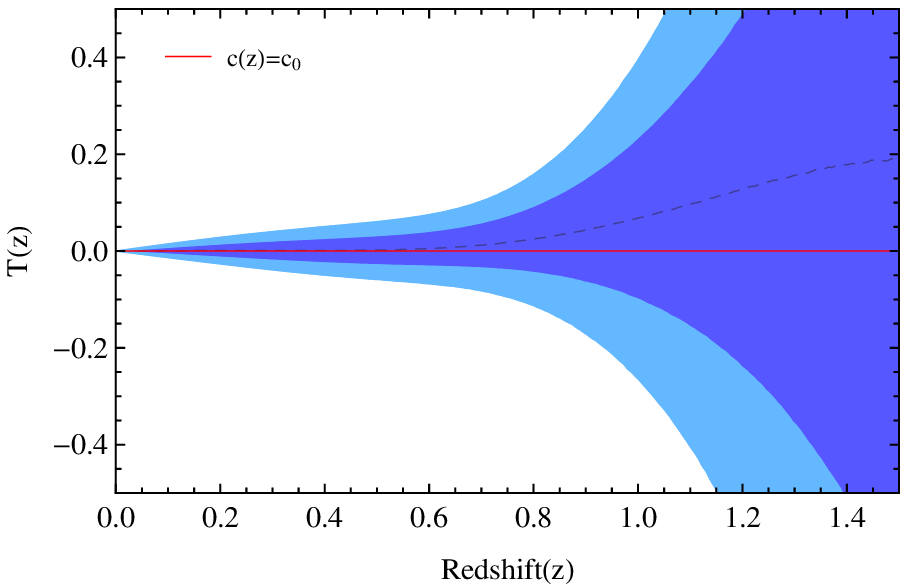}}\quad
\subfloat{
\includegraphics[width=0.25\textwidth]{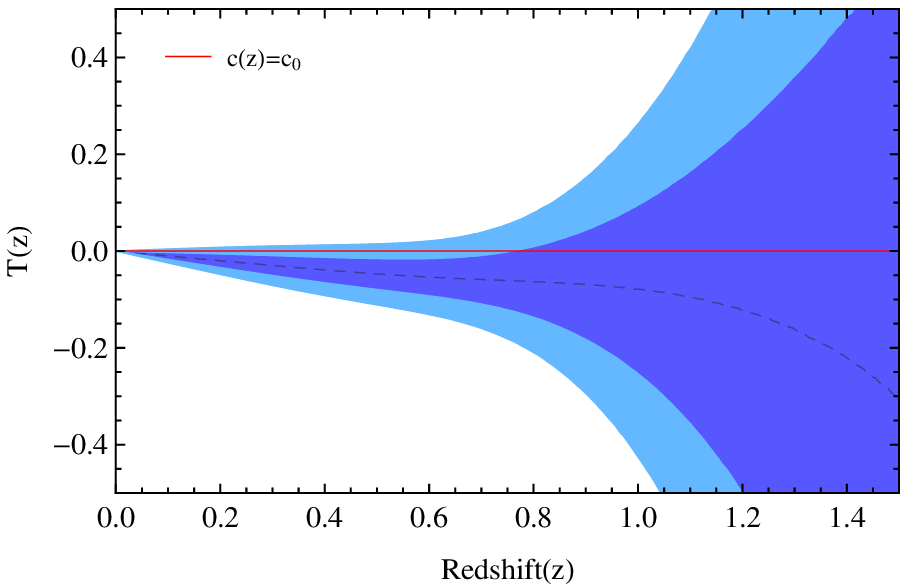}}\quad
\subfloat{
\includegraphics[width=0.25\textwidth]{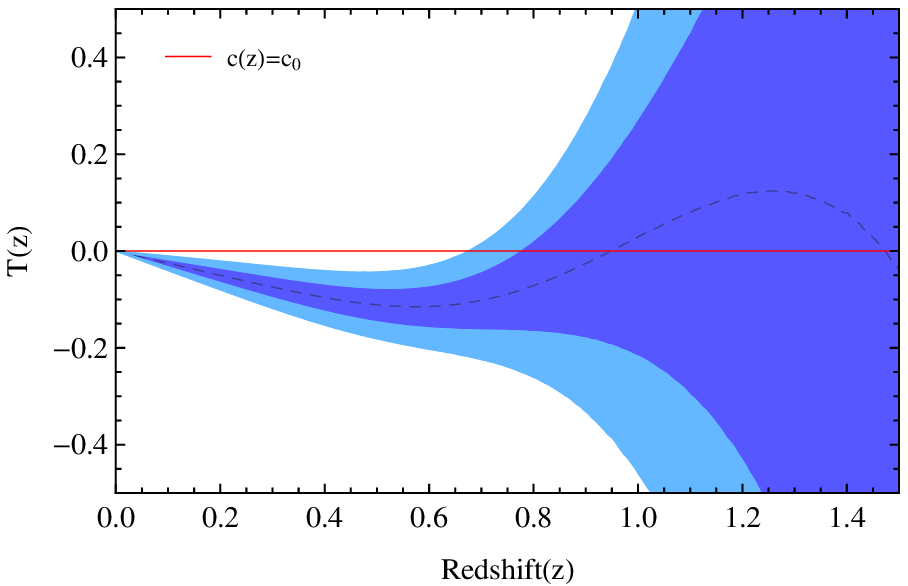}}\\
\subfloat{
\includegraphics[width=0.25\textwidth]{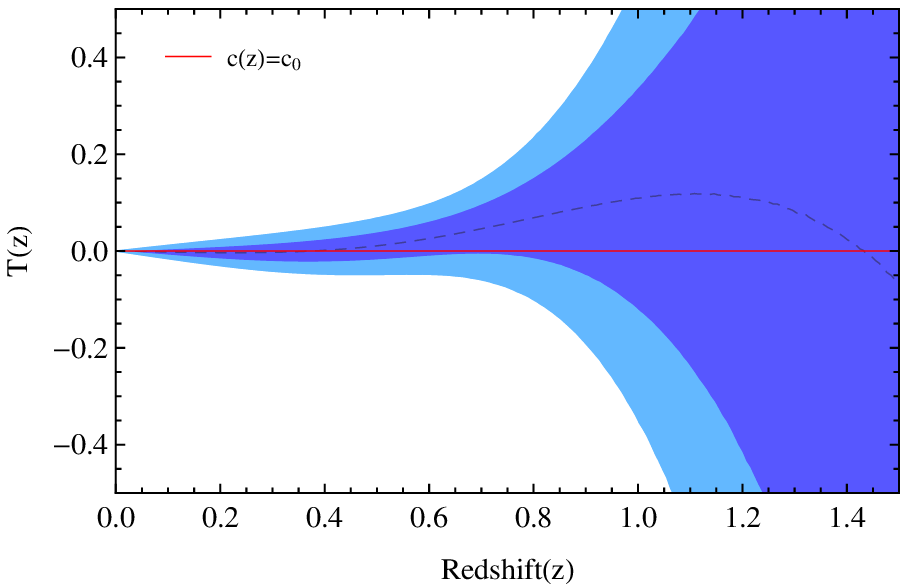}}\quad
\subfloat{
\includegraphics[width=0.25\textwidth]{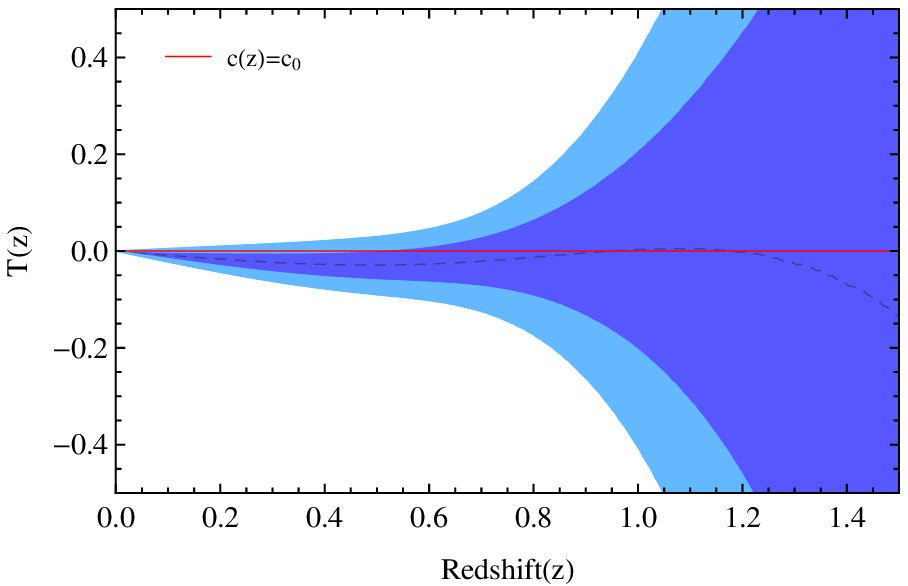}}\quad
\subfloat{
\includegraphics[width=0.25\textwidth]{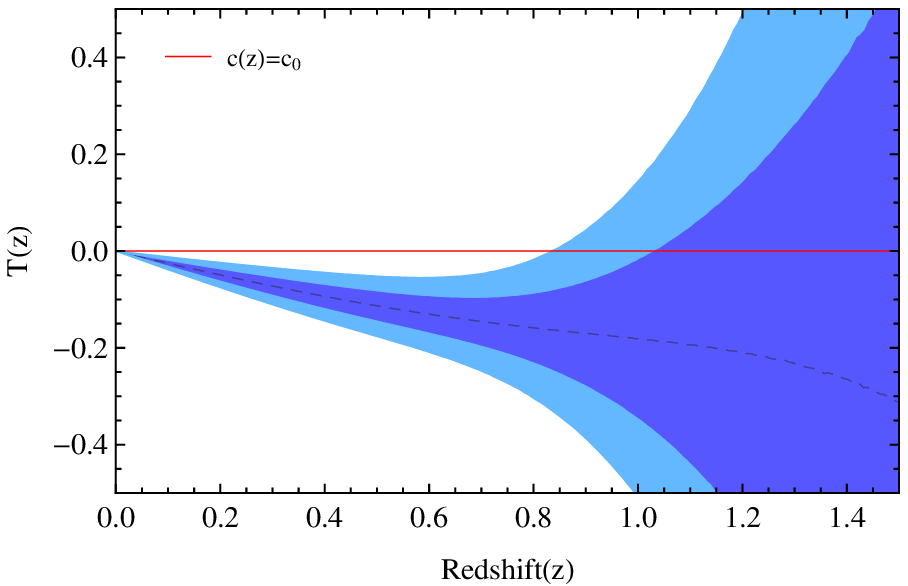}}\\
\subfloat{
\includegraphics[width=0.25\textwidth]{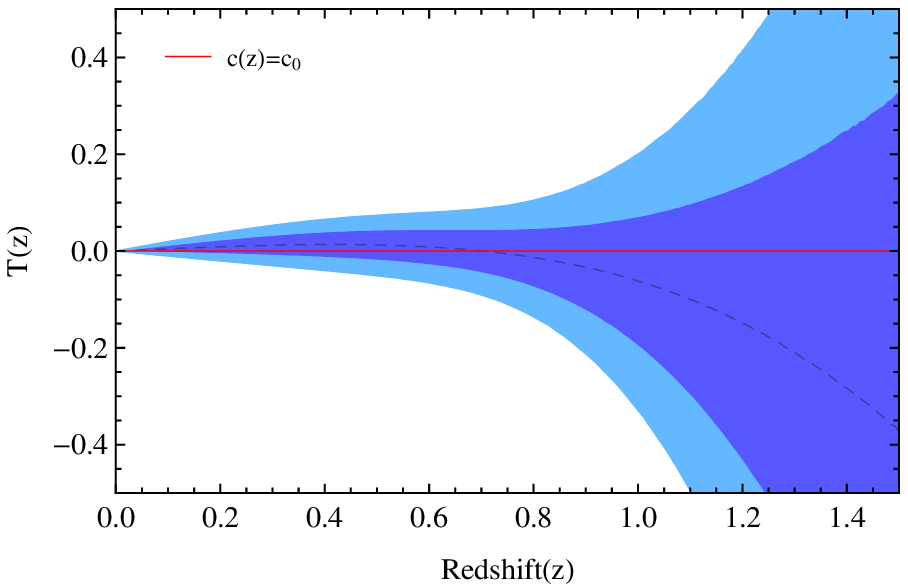}}\quad
\subfloat{
\includegraphics[width=0.25\textwidth]{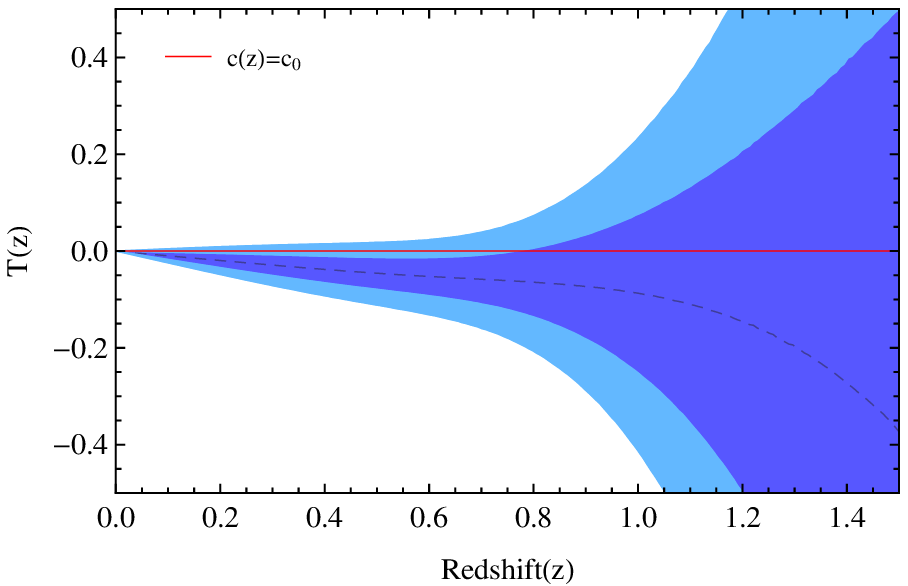}}\quad
\subfloat{
\includegraphics[width=0.25\textwidth]{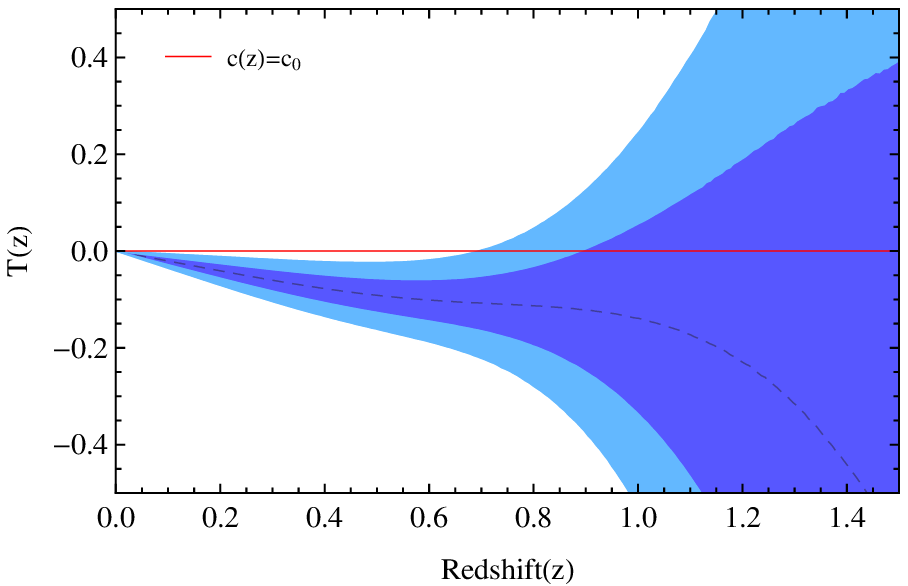}}\quad
\caption{\label{fig:mock}Reconstructions of $T(z)$ for different models. From top to bottom, $\Omega_K=0,0.16,-0.16$, respectively. And from left to right, $\Delta c = 0$, $\Delta c /c_0 \sim 1\%$, $\Delta c /c_0 \sim 2\%$, respectively. The shaded blue regions are the $68\%$ and $95\%$ C.L. for the reconstruction.}
\label{fig:mock}
\end{figure}

\begin{figure}
\centering
\includegraphics[width=0.45\textwidth]{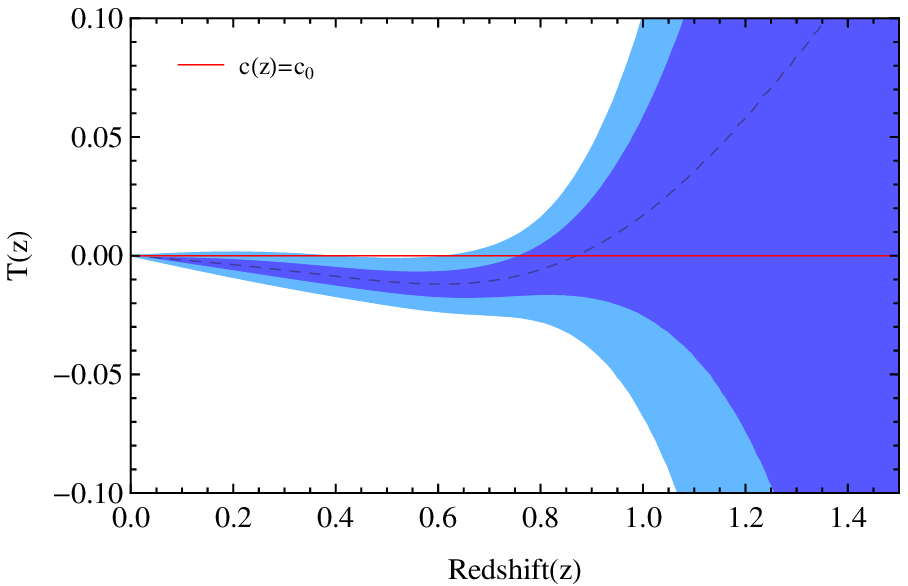}
\caption{Reconstructions of $T(z)$ for $\Delta c /c_0 \sim 0.1\%$. The error is one-tenth of the original mock data. The shaded blue regions are the $68\%$ and $95\%$ C.L. of the reconstruction. The red line corresponds to $c_0$.}
\label{fig:mockc01}
\end{figure}

\subsection{Mock data}

In order to evaluate the constraint on the null test from future observational experiments and its ability to probe the small variation of the speed of light from a constant, we create mock data to test the ability. Following~\cite{Salzano:2014lra,Salzano:2015mxk}, we choose a general theoretically-motivated expression given in~\cite{Magueijo:2003gj} as
\begin{equation}
c(a) \propto {c_0}{(1 + \frac{a}{{{a_c}}})^n},
\label{equa:vsl}
\end{equation}
where $a\equiv 1/(1+z)$ is the scale factor. Note that here we should add a normalization factor to make $c(a=1)=c_0$. Since our model-independent method can probe the constancy of the speed of light at every redshift, we can focus on the redshift range where the high quality of the data sets lie, i.e. $z\in [0.2,0.6]$.

We firstly simulate a data set of 128 points for $E(z)$. Adopting the methodology in~\cite{Ma:2010mr}, we use the errors of current observational data. We draw the error from a Gaussian distribution: $\sigma_{E} \sim \mathcal{N}(\bar{\sigma},\epsilon)$ with $\bar{\sigma} =(\sigma_+ + \sigma_- )/2$ and $\epsilon = (\sigma_+ - \sigma_- )/4$, where $\sigma_+$ and $\sigma_-$ are the two straight lines bound the uncertainties $\sigma(z)$ of the observational $E(z)$ data from above and below, respectively. Then $E(z)_{sim}$ is sampled from the Gaussian distribution $E(z)_{sim} \sim \mathcal{N}(E(z)_{fid},\sigma_{E})$, where $E(z)_{fid}$ is the theoretical value of the fiducial model.

For simulated $D(z)$ data, we create mock data sets of future SNeIa according to the Dark Energy Survey (DES)~\cite{Bernstein:2011zf}. The DES is expected to obtain high quality light curves for about 4000 SNeIas from $z=0.05$ to $z=1.2$. From Table $14$ in~\cite{Bernstein:2011zf} we can calculate the errors in $D$ and $\sigma_{D}$ and the corresponding numbers of SNeIa for each redshift bin. At every redshift point $z$, $D(z)_{sim}$ is sampled from the normal distribution $D(z)_{sim} \sim \mathcal{N}(D(z)_{fid},\sigma_{D})$.

In our analysis, we produce mock data based on three different cosmological models: (1) $\Delta c = 0$, the baseline $\Lambda$CDM model; (2) ${a_c}=0.05$,  $n=-0.04$, $\Delta c /c_0 \sim 0.5\%-1.5\%$ at $z\in [0.2,0.6]$; (3) ${a_c}=0.05$,  $n=-0.09$, $\Delta c /c_0 \sim 1.5\%-3\%$ at $z\in [0.2,0.6]$. Additionally, in these three models we set $\Omega_K = 0$, $+0.16$ and $-0.16$, respectively, which is significantly large to be detected by current data sets using a model-independent method proposed in~\cite{Cai:2015pia}. If the null-tests with different values of the cosmic curvature have the same results, it implies that we  have indeed dodged the cosmic curvature. Here we emphasize that in the case of non-vanishing cosmic curvature, the corresponding Friedmann equation is
\begin{equation}
{H^2} = \frac{{8\pi G}}{3}\rho  - K{(1 + z)^2}{c^2}.
\label{equa:H}
\end{equation}
As illustrated in Fig~\ref{fig:mock}, we can distinguish these three models correctly. $\Delta c /c_0 \sim 1\%$ can be detected at $\sim 1.5\sigma$ C.L. and $\Delta c /c_0 \sim 2\%$ or lager can be detected at $\sim 3\sigma$ C.L. The results with different values of the cosmic curvature are consistent, which indicates that we indeed dodged the cosmic curvature as discussed in Sec.~\ref{sec:tb}.

Comparing our results to those obtained in the papers~\cite{Salzano:2014lra,Salzano:2015mxk}, we find that DES can not provide better improvement in detecting variation of the speed of light for the $\Delta c /c_0 \sim 1\%$ case. This is because dodging the cosmic curvature introduces the second derivatives which will lead to larger errors on the results. Though we have this disadvantage, our method has almost the same ability to detect the $\Delta c /c_0 \sim 1\%$ deviation. If we want to detect smaller deviation from $c_0$, we have to use the data sets with higher quality, {\it i.e.}, bigger number and smaller errors of data. Following~\cite{Salzano:2015mxk} we also produce the cosmic chronometers $H(z)$ and SNeIa data from DES, but with errors ten times smaller than the expected DES ones. The result is shown in Fig.~\ref{fig:mockc01}. We can see that reducing the errors to one-tenth of the expected DES ones will make it possible to detect $\Delta c /c_0 \sim 0.1\%$ at $2\sigma$ C.L.

\section{Conclusions and discussions \label{sec:discussion}}

In this paper, we develop a new method to probe the constancy of the speed of light. By dodging the cosmic curvature, we can directly test the speed of light without assuming any cosmological model, which should be  more natural than the method proposed in~\cite{Salzano:2014lra}. Furthermore, we can test $c(z)$ at every redshift covered by  the data sets, so that we can use real observational data and it is not limited in the constrained region of the redshifts. We use $H(z)$ data from the cosmic chronometers and BAO, and then combine them with SNIa Union 2.1 and the most recent JLA to give a null test of the speed of light. The result indicates that there is no signal of deviation from $c_0$. What we are more concerned about is how small variation of the speed of light we can detect using higher quality data sets in the future. For this purpose, we create mock data sets based on three fiducial models. We find that using the simulated data sets whose errors are obtained from cosmic chronometers and DES, we can detect $\Delta c /c_0 \sim 1\%$ at $\sim 1.5\sigma$ C.L. and $\Delta c /c_0 \sim 2\%$ at $\sim 3\sigma$ C.L. If we improve the quality of the data, i.e, set the errors to be one-tenth of the expected DES ones, we can easily detect a $\Delta c /c_0 \sim 0.1\%$ variation at $2\sigma$ C.L. The luminosity distance containing more a factor $\hat c(z)$ outside the integral makes it can detect more information of the speed of light.

Dodging the cosmic curvature is crucially important when we want to detect a smaller variation of the speed of light because even a very small cosmic curvature can influence the results due to the strong degeneracy between them. Our new method developed in this paper overcomes such a problem. Although our method has no advantage in the accuracy compared to the one in~\cite{Salzano:2014lra} using current data sets, it provides a model-independent way to detect variation of the speed of light with more precision when the quality of the data sets can be improved in the future. Finally we mention that according to (\ref{equa:cz}), in principle, we can
reconstruct the speed of light, once one knows $A(z)$, $B(z)$, and $M(z)$ from the observational data.

In fact, changing the speed of light at a redshift $z$ leads to a variation in the fine structure constant $\alpha\equiv e^2/(\hbar c)$,
where $e$ is the electron charge and $\hbar$ the reduced Planck constant~\cite{Dzuba:1999zz}.
If the parameters $e$ and $\hbar$ involved in its definition are assumed to be constant, it is easy to get $\Delta c/c_0=-\Delta \alpha/\alpha$,
which impacts the measurement of the redshift $z$ of the object.
Recent analysis in Ref.~\cite{Rahmani:2013pva} implies that the variation in the fine structure constant is very small,
at least $\Delta \alpha/\alpha < 10^{-4}$.
Therefore, we can ignore the effect of the variation in $\alpha$ on the measurement of the redshift.
In principle, a large variation in the speed of light can still be compatible with such orders of magnitudes
if the other parameters are allowed to vary.

It is known that the distance-duality between the luminosity distance and the angular-diameter distance
is violated by non-conservation of photon number~\cite{Bassett:2003vu,Kunz:2004ry}.
Of course, such a violation of the distance-duality results in the deviation of $T(z)$ in Eq.~\eqref{equa:ctest} from zero.
We have to emphasize that only the change of the speed of light has been considered in this paper.


\acknowledgments
This work is supported by the Strategic Priority Research Program of the Chinese Academy of Sciences, Grant No.XDB09000000.
Z.K.G is supported by the National Natural Science Foundation of China Grants No.11575272 and No.11335012.


\end{document}